\documentclass{article}
\usepackage{amsmath}
\usepackage{amsfonts}
\usepackage{amssymb}
\advance\topmargin by -5mm \advance\textheight by 10mm

\newcommand{\roton}{CARP}
\newcommand{\vs}{\ensuremath{v_\text{s}}}
\author{Lev A.\,Melnikovsky\thanks{E-mail: leva@kapitza.ras.ru}\\
\\
\textit{P.L.\,Kapitza Institute for Physical Problems}\\
\textit{Russian Academy of Sciences, Moscow, Russia}}
\title{Josephson Effect in Coherent Roton Aggregates}
\date{}
\begin{document}
\maketitle

\begin{abstract}
A microwave electromagnetic field can excite a coherent roton aggregate
in liquid helium around a dielectric resonator. We show that multiple
coherent aggregates are excited simultaneously and predict a Josephson
effect between them. The superfluid velocity acts as a ``voltage across
the weak link'' in superconducting Josephson junctions. A comparison
with existing experimental data is made.
\end{abstract}


It was shown earlier \cite{rpr1,rpr2}, that a Coherent Aggregate of
Roton Pairs (\roton) in liquid helium-4 around a dielectric
resonator can be excited by an electromagnetic field
in the microwave range.
Experimentally, the \roton\ build-up manifests \cite{ryba1} as
an ultra-narrow peak in the resonator loss at the frequency of
$f_0=\Delta/(2\pi\hbar)\sim 180\,\text{GHz}$, where $\Delta$ is the
roton energy gap.  Coupling of the microwave radiation to the
\roton\ is due to the dependence of individual roton energy
$\varepsilon$ on the electric field $\mathbf{E}(t)$:
\begin{equation*}
\delta\varepsilon \sim \alpha\frac{E^2}{2},
\end{equation*}
where $\alpha$ is the roton polarizability. Elementary process of such
parametric excitation is the transformation of two photons into two rotons.
Energy of the photon pair $4\pi\hbar f$ in the initial state is equal to the energy of the
roton pair $2\varepsilon = 2\Delta$.
The photon momentum was earlier \cite{rpr2} neglected
because it is much smaller than that of a roton.
This momentum is in fact responsible for an interesting effect
described below.

Electromagnetic field in a whispering gallery disk resonator is a
superposition of two traveling modes --- a counterclockwise wave and a
clockwise wave. The azimuthal projection of the wave vector for
these modes is
\begin{equation*}
k=\pm\frac{N}{R},
\end{equation*}
where $R$ is the resonator radius and $N$ is the mode number. The
azimuthal momentum of the roton pair may therefore take one of three
possible values
\begin{equation*}
p_-=-\frac{2\hbar N}{R}, \quad
p_0=0,\quad \text{or}\quad
p_+=\frac{2\hbar N}{R}.
\end{equation*}
This means that actually three \roton s are excited by the parametric
resonance rather than just one. Suppose a weak interaction exists
between these coherent particle reservoirs. The internal Josephson
effect is to be expected in such system. The Josephson currents between
the \roton s would be determined by the phase differences between them.

Roton pairs of different azimuthal momenta in a motionless liquid have
equal energy $2\Delta$. This degeneracy may be removed by a superfluid velocity \vs.
Imagine an axially symmetric vortex superflow tangential to the disk
circumference. The energy of the roton pair then becomes
\begin{equation*}
2\varepsilon_{\{-,0,+\}}=2\Delta + \vs p_{\{-,0,+\}}.
\end{equation*}
Note, that \vs\ here is not arbitrary, it depends on the number of 
quantized vortices $n$ pinned by the resonator:
\begin{equation*}
\vs=\frac{n\hbar}{R m_\text{He}}.
\end{equation*}

The time-dependant parts of the phase differences are simply $\Omega_n t$ and
$2 \Omega_n t$, where
\begin{equation}
\label{step}
\Omega_n=\frac{\vs p_+}{\hbar} =
\frac{2\hbar N n}{R^2 m_\text{He}} = 2\pi n \cdot 0.0043\,\text{Hz}.
\end{equation}
Here $R=9.5\,\text{mm}$ and $N=78$ (see \cite{ryba2}). With the
frequency $\Omega_n$ both the Josephson currents and \roton s population
oscillate. The latter is directly probed in microwave experiments.
Indeed, the step-like behavior and the low frequency modulation of the
resonator loss have been observed \cite{ryba2}. In this experiment the
superfluid circulation around the disk was generated by two ``heat
guns'' and the modulation amplitude changed stepwise with continuous
increase of the gun power. These steps can be attributed to the
circulation quantization. It should be possible to extract the
oscillation frequency step corresponding to one vortex quantum.
Unfortunately, available data \cite{ryba2} leave certain
amount of uncertainty about exact value of this step, it is probably
confined within the range $0.002\,\text{Hz} - 0.04\,\text{Hz}$ in good
agreement with~\eqref{step}.

It is possible that small Josephson oscillation are masked in experiment
by some spurious beat-frequency interference between electromagnetic
waves absorbed by \roton s of different momenta. The lowest beat frequency
is $\Omega_n/2$.

I thank A.F.\,Andreev, V.I.\,Marchenko, A.S.\,Rybalko, and A.I.\,Smirnov for fruitful discussions.
This work was supported in parts by RF president program NSh-4889.2012.2 and
RFBR grants 13-02-00912 and 13-02-90494.

\end{document}